\documentclass[aps,prl,twocolumn,groupedaddress]{revtex4-2}
\usepackage{graphicx}
\usepackage{lipsum}
\usepackage{float}
\usepackage{babel}
\usepackage[utf8]{inputenc}
\usepackage{color}
\usepackage{dcolumn}
\usepackage{bm}

\begin{document}

\title{Sedimentation and Levitation of Catalytic Active Colloids}
\author{V. Carrasco-Fadanelli}
\email{carrasv@hhu.de}
\affiliation{Institute of Experimental Colloidal Physics, Department of Physics, Heinrich-Heine-Universität Düsseldorf, 40225 Düsseldorf, Germany}
\author{I. Buttinoni}
\affiliation{Institute of Experimental Colloidal Physics, Department of Physics, Heinrich-Heine-Universität Düsseldorf, 40225 Düsseldorf, Germany}

\date{\today}

\begin{abstract}

Gravitational effects in colloidal suspensions can be easily turned off by matching the density of the solid microparticles with the one of the surrounding fluid. By studying the motion of catalytic microswimmers with tunable buoyant weight, we show that this strategy cannot be adopted for active colloidal suspensions. If the average buoyant weight decreases, pronounced accumulation at the top wall of a sample cell is observed due to a counter-alignment of the swimming velocity with the gravitational field. Even when the particles reach a flat wall, gravitational torques still determine the properties of the quasi two-dimensional active motion. Our results highlight the subtle role of gravity in active systems.    
\end{abstract}

\maketitle

Active Brownian particles are a new generation of colloids which undergo self-propulsion rather than being in thermal equilibrium with their liquid environment \cite{RevModPhys.88.045006,Ramaswamy2010,B918598D}. They have been receiving much attention as model systems for living matter \cite{you2018intelligent}, microvehicles for the technological applications (e.g. in the realm of drug delivery) \cite{katuri2017designing,xu2020self}, and building blocks for the design of new materials with self-sustaining properties \cite{C7CS00461C,PhysRevE.91.012134,palacci2013living,PhysRevLett.110.238301}. 

Many of the existing synthetic microswimmers take advantage of interfacial phenomena such as phoresis and osmosis to achieve self-propulsion \cite{anderson1989colloid}. In short, patchy colloids can generate local asymmetric conditions (e.g., thermal or chemical gradients) that, in turn, give rise to slip flows and directed motion \cite{aubret2017eppur}. The most widespread example -- also of main interest for the present study -- is that of spherical particles half-coated with a layer of platinum ($\rm Pt$) and suspended in hydrogen-peroxide ($\rm H_2O_2$) enriched solutions \cite{howse2007self,dietrich2017two,ketzetzi2020slip,brown2014ionic,PhysRevLett.105.088304}. The platinum cap acts as a catalyst and locally decomposes the `fuel' into $\rm H_2O$ and $\rm O_2$ leading to autophoretic swimming velocities that increase with the bulk concentration of hydrogen peroxide, $\rm [H_2O_2]_{\infty}$. Importantly, in the absence of confinements or external fields, the velocity vector is always oriented along the axis linking the poles of the two hemispheres and rotates together with the particle.  

To date, experiments with synthetic active particles are almost entirely limited to (quasi) two-dimensional (2D) systems, with only few exceptions \cite{doi:10.1063/1.5124895,sakai2020active}. Self-propelling colloids heavier than the solvent sediment towards the bottom of the container (\textit{positive gravitaxis}), although fast bottom-heavy \cite{campbell2013gravitaxis,doi:10.1063/1.4998605} and asymmetric \cite{ten2014gravitaxis} active particles may also levitate (\textit{negative gravitaxis}) due to a counter-alignment with the gravitational field. Positive and negative gravitaxis have been studied as a function of the swimming velocity and size \cite{campbell2013gravitaxis,doi:10.1063/1.4998605}, but we are not aware of experimental studies addressing specifically the role of the buoyant weight -- which is paramount to achieve zero effective gravity -- of the self-propelling particles. Likewise, existing studies of catalytic self-propulsion near flat confinements only address the case of `heavy' colloids; they unanimously suggest that, in the presence of underlying substrates, active colloids exhibit a steady motion with a fixed distance from the substrate and fixed orientation with respect to vertical axis \cite{dietrich2017two,das2015boundaries,ketzetzi2020slip,ketzetzi2020diffusion,das2020floor,uspal2015self}.   

In this Letter, we fine-tune the buoyant weight of catalytic microswimmers and study how it influences the active motion in bulk and near solid surfaces. Buoyancy plays a key role in the direction of particle migration, {\sl i.e.} whether the colloids sink to the bottom substrate, levitate towards the top wall or remain in the bulk of the sample cell. Experiments and Brownian dynamics simulations reveal that, if the average buoyant weight becomes close to zero, accumulation takes place almost entirely at the top wall due to a `residual' gravitational torque. Once the particles reach a substrate, they undergo an oscillatory motion (in the direction of gravity) with a frequency and amplitude that depend on both the swimming velocity and buoyant weight.  

\begin{figure}[t]
	\includegraphics[scale=0.45]{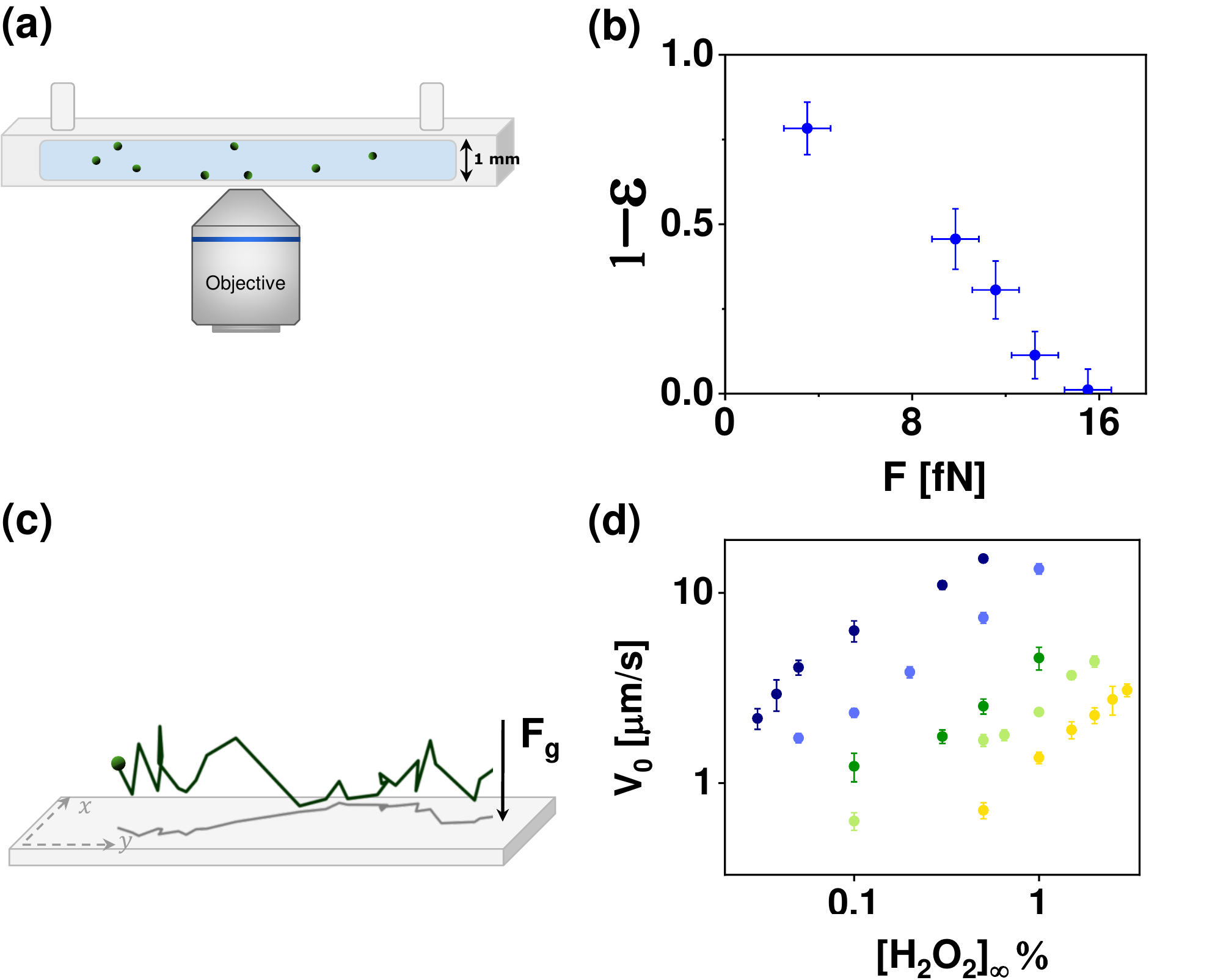}
\caption{(a) Sketch of the experimental setup. A quartz cell is filled with a suspension of active particles in $\rm H_2O$-$\rm D_2O$-$\rm H_2O_2$ solutions and imaged from below using an inverted microscope. (b) Fraction of `passive' particles ($[\rm H_{2}O_{2}]_{\infty}=0$ \%) in the bulk of the cell, $1-\varepsilon$, one hour after injection, plotted as a function of the buoyant weight, $F$. The data are in agreement with simulations of `passive' Brownian particles (not shown). (c) Experimental trajectories in 2D (gray line) and 3D (green line) of an active particle close to the bottom substrate. Gravity points in the \textit{z}-direction. The mean velocity in the $xy$-plane (plane parallel to the substrate) is obtained by fitting the 2D-MSD. (d) $V_{0}$ {\sl vs.} $\rm [H_2O_2]_{\infty}$ for five different buoyant weights. From blue to yellow: $F =15.52$ $\rm fN$, $F = 13.25$ $\rm fN$, $F = 11.57$ $\rm fN$, $F = 9.85$ $\rm fN$, $F = 3.51$ $\rm fN$.}\label{fig1}
\end{figure}

The patchy particles consist of spherical fluorescent polystyrene ($\rm PS$) beads (Microparticles GmbH, radius $R= 1.4$ $\mu {\rm m}$) coated on one of their hemispheres with a $\rm Pt$-layer of thickness $\lambda = 4 \pm 0.5$ $\rm nm$ \cite{SupplementalMaterial}. The particles are homogeneously dispersed in $\rm H_2O$-$\rm D_2O$-$\rm H_2O_2$ solutions (particle volume fraction, $\phi \approx 0.005 \%$ $\rm \rm v/\rm v$) and pipetted in a $35$ $\times$ $9$  $\times$ $1$ $\rm mm^3$ quartz cell (Hellma Analytics, see sketch in Fig.~\ref{fig1}(a)), whose inlet and outlet are later sealed. Each colloid is subjected to two forces: a gravitational weight, which is the sum of the weights of the $\rm PS$-sphere and $\rm Pt$-coating, {\sl i.e.}
 
 \begin{equation}
  	\textbf{F}_{\rm g} = \left[ \frac{4}{3}\pi R^3 \rho_{\rm ps}  + 2\pi \lambda R^2 \rho_{\rm pt} \right] \textbf{g},
 \end{equation}\label{GravitationalForce}
 
 \noindent and a buoyant force
 
  \begin{equation}
 	\textbf{F}_{\rm b} = - \frac{4}{3}\pi R^3 \rho_{\rm s} \textbf{g}.
 \end{equation}\label{GravitationalForce} 
 
\noindent Here, $\textbf{g}$ is the gravity acceleration (pointing towards the bottom wall), and $\rho_{\rm ps} = 1.05$ $\rm g/cm^3$, $\rho_{\rm pt} = 21.45$ $\rm g/cm^3$ and $\rho_{\rm s}$ are the densities of the polystyrene particles, platinum cap and solvent, respectively. In our experiments, the coating contributes to approximately 10 \% of the total gravitational weight. The addition of $\rm D_2O$ allows us to vary $\rho_{\rm s}$ and, as such, achieve a buoyant force that almost entirely counteracts $\textbf{F}_{\rm g}$ ($\rm [D_2O] = 100$ \%). Within the experimental timescale of one hour, this leads to nearly weightless colloids (buoyant weight, $\textbf{F}=\textbf{F}_{\rm g}+\textbf{F}_{\rm b} \simeq 0$), as shown by the large fraction ($1-\varepsilon$) of particles staying in the bulk (Figure~\ref{fig1}(b); note, however, that at longer times the particles will eventually sediment). We shall see that this conclusion is valid only in the absence of activity ($\rm [H_2O_2]_{\infty} = 0$). 
 
Soon after injection, catalytic active colloids undergo self-propulsion in bulk with the $\rm PS$-hemispheres heading; some go to the top wall of the cell, some to the bottom wall and others remain in the bulk. We first record images of active colloids swimming on the bottom substrate at $5$ frames per second by means of an inverted fluorescence microscope and a CMOS camera, and extract the $xy$-trajectories (grey trajectory in Fig.~\ref{fig1}(c)) using Matlab scripts. The mean swimming velocity in $xy$, which we denote as $V_{\rm 0}$, is obtained by fitting mean squared displacement (MSD) of approximately $15$ particles \cite{SupplementalMaterial}. As opposed to all existing experiments, $V_{\rm 0}$ is not only a function of $\rm [H_2O_2]_{\infty}$, but also the ratio between $\rm H_2O$ and $\rm D_2O$, possibly due to the higher stability of deuterium peroxide \cite{giguere1940heterogeneous}. As the concentration of $\rm D_2O$ increases, more hydrogen peroxide is needed to obtain the same swimming velocity, as illustrated by the different colours (from dark blue, $\rm [D_2O] = 0 \%$, to yellow, $\rm [D_2O] = 100 \%$) of the experimental data in Fig.~\ref{fig1}(d). We often observe a linear growth of $V_{\rm 0}$ as a function of the fuel concentration, although a more complex dependence is reported in $\rm H_2O$-$\rm H_2O_2$ mixtures for small values of $\rm [H_2O_2]_{\infty}$ (a plot of the data for $\rm [D_2O] = 0 \%$ and a broader range of  $\rm [H_2O_2]_{\infty}$ is given in the Supplementary Material \cite{SupplementalMaterial}). $V_{0}$ is approximately the same regardless of the cell wall (bottom {\sl vs.} top) they are close to.   
 
\begin{figure}[t]
 	\includegraphics[scale=0.45]{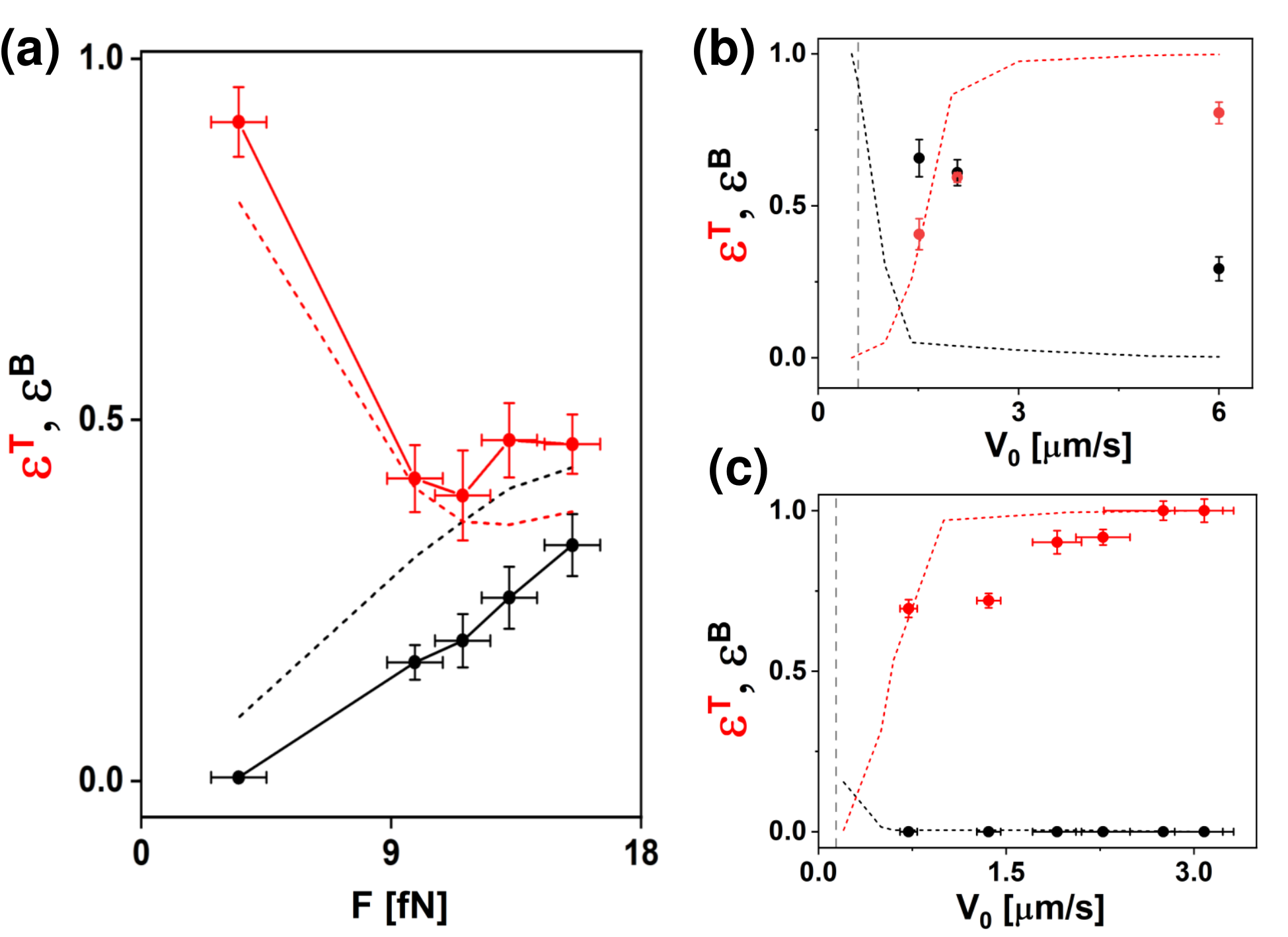}
 	\caption{(a) Ratio of particles found at the bottom ($\varepsilon^{\rm B}$, black) and top ($\varepsilon^{\rm T}$, red) walls of the cell after one hour, plotted as a function of the buoyant weight $F$, for $V_{0} = 2$ $\mu \rm m/s$. (b-c) $\varepsilon^{\rm T}$ and $\varepsilon^{\rm B}$ {\sl vs.} $V_{0}$ for (b) $F=15.52$ $\rm fN$ ($\rm [D_2O] = 0 \%$) and (c) for $F=3.51$ $\rm fN$ ($\rm [D_2O] = 100 \%$). The dashed vertical lines denote the velocity at which the buoyant weight is equal to the Stokes drag, $6 \pi \eta R V_0$. The dotted lines are numerical simulations, as described in the main text.}\label{fig2} 
\end{figure}

By controlling $\rm [D_2O]$ and $\rm [H_2O_2]_{\infty}$, we look at the ratio of particles that accumulate at the top ($\varepsilon^{\rm T}$) and bottom ($\varepsilon^{\rm B}$) walls as a function of $F$ ($V_{0}$), while keeping $V_{0}$ ($F$) constant ($\varepsilon=\varepsilon^{\rm B}+\varepsilon^{\rm T}$). Figure~\ref{fig2}(a) shows $\varepsilon^{\rm T}$ (red data) and $\varepsilon^{\rm B}$ (black data) for particles swimming at $\rm V_{0}=2$ $\rm \mu m/s$. As $F$ increases, the particles sediment to the bottom substrate, albeit at a slower rate compared to `passive' Brownian microbeads ($V_{0}=0$). One reason for the slower sedimentation rate lies in the larger effective thermal energy of active colloids with long-time diffusion coefficient \cite{RevModPhys.88.045006},

\begin{equation}
	D_{\rm eff} = \frac{k_{\rm B}T_{\rm eff}}{6 \pi \eta R} = D_0 + \frac{1}{2} V_0^2 \tau_{\rm R},
\end{equation}

\noindent where $k_{\rm B}$ is the Boltzmann constant, $\eta$ is the solvent viscosity,  $T_{\rm eff}$ is an effective temperature, $D_0=(k_{\rm B}T)/(6 \pi \eta R)$ is the translational diffusion coefficient, and $\tau_{\rm R} = (8 \pi \eta R^3)/(k_{\rm B}T)$ is the timescale of rotational diffusion. Hence, active colloids are `hotter' than their passive counterparts, {\sl i.e.} their sedimentation lengths, $l_{\rm g}= k_{\rm B}T_{\rm eff}/F$, are significantly larger (the values are reported in the Supplementary Material \cite{SupplementalMaterial}). On the other hand, when $F \rightarrow 0$, accumulation is observed almost entirely at the top wall of the cell; this behaviour cannot be rationalized neither by taking into account an effective diffusivity, which does not have a preferential direction, nor the persistence length, $L=V_{0}\tau_{\rm R}$, which is much smaller than the cell's height. In Fig.~\ref{fig2}(b-c), we plot $\varepsilon^{\rm B}$ and $\varepsilon^{\rm T}$ as a function of the swimming velocity for the limit cases of pure $\rm H_2O$-$\rm H_2O_2$ ($F=15.52$ $\rm fN$) and pure $\rm D_2O$-$\rm H_2O_2$ ($F=3.51$ $\rm fN$) mixtures. In water (Fig.~\ref{fig2}(b)), slow particles sediment towards the bottom substrate and fast swimmers migrate to the top wall. The black and red data show a crossover at $V_{0} \simeq 2$ $\mu \rm m /s$, which is slightly larger than the velocity at which the buoyant weight is equal to the Stokes drag (dashed vertical line). In contrast, pure deuterium oxide entirely favours levitation within our experimental range (Fig.~\ref{fig2}(c)).

\begin{figure}[t]
	\includegraphics[scale=0.5]{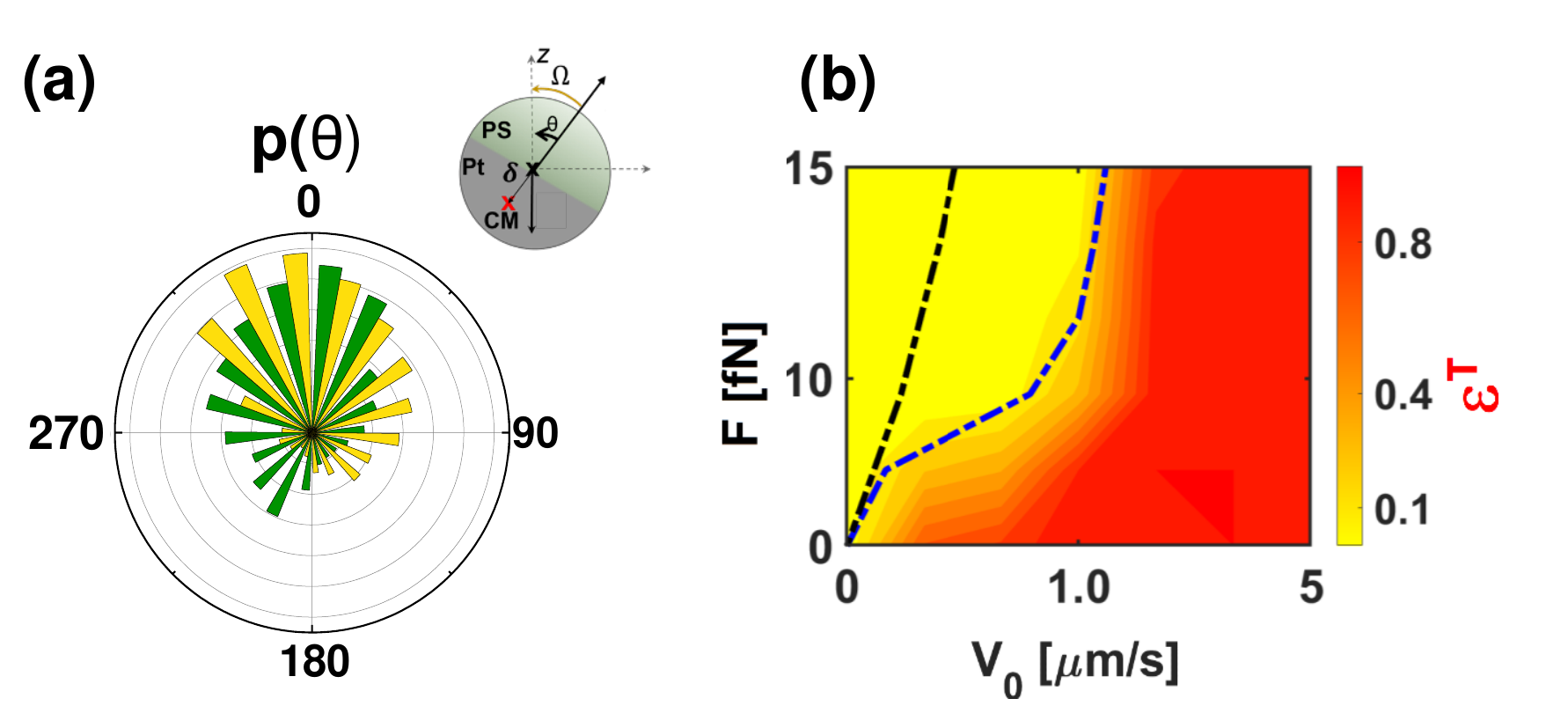}
	\caption{(a) Normalized distribution $p(\theta)$ of the particle's orientation, $\theta$, from numerical simulations at $V_{0}=2$ $\rm \mu m/s$ for $F =15.52$ $\rm fN$ (yellow) and $F =0$ $\rm fN$ (green). $\theta=0$ corresponds to particles oriented antiparallel to gravity. Inset: Schematic representation of the model active particles used in simulations. As the particle is subjected to gravitational force, it experiences a torque $\bm{\Omega}$ that aligns the swimming velocity with the $z$-axis. (b) Numerical results for the fraction of active particles at the top surface ($\varepsilon^{\rm T}$) as a function of $F$ and $V_{0}$. The dashed blue line denotes the critical swimming velocity, $V_{0}^{*}$, for which $\varepsilon^{\rm T}=\varepsilon^{\rm B}$.}\label{fig3} 
\end{figure}

The experimental data in Fig.~\ref{fig2} are fully rationalized by considering that the $\rm Pt$-coated microparticles are bottom-heavy \cite{campbell2013gravitaxis,PhysRevLett.107.058301,Stark2016}, {\sl i.e.} they are subjected to a torque that favours a swimming direction against the gravitational field (\textit{negative gravitaxis}). We employ overdamped Brownian dynamics simulations \cite{Callegari2019} where the patchy microswimmers are modelled as active Brownian disks equipped with translational velocity $V_{0}$ that points towards the pole of the uncoated hemisphere and rotates in the $xz$-plane. Thermal fluctuations are included in both the translational and rotational motion. The disks are subjected to a buoyant weight $\textbf{F}$ and experience a torque $\bm{\Omega}$ due to a point-like weight $\textbf{F}_{\rm Pt}=2 \pi \lambda R^2 \rho_{\rm Pt} \textbf{g}$ ($\rho_{\rm Pt} = 21.45$ $\rm g/cm^3$) located on the axis of the velocity vector at a distance $\delta \simeq 0.3 R$ from the geometric center of the particle, as schematically illustrated in the inset of Fig.~\ref{fig3}(a). $\bm{\Omega}$ tends to align the velocity vector with the $z$-axis (Fig.~\ref{fig3}(a)). Importantly, the torque still exists even if $F \simeq 0$ and is in fact approximately constant for all values of $\textbf{F}$ because platinum is much heavier than the solvent for any concentration of deuterium oxide. This is in stark contrast with existing studies of gravitactic bottom-heavy swimmers, in which the weight is varied using particles of different size \cite{campbell2013gravitaxis,doi:10.1063/1.4998605}. The governing equations used for the simulations and a detailed description of the gravitational torque experienced by the particles are in the Supplementary Material \cite{SupplementalMaterial}.

A particle is initially placed at a random height and self-propels in a rectangular box of height $1$ $\rm mm$ until it reaches the top or bottom boundary, at which point the simulation ends and a contribution towards $\varepsilon^{\rm T}$ or $\varepsilon^{\rm B}$ is counted. If the particle has not reached any wall after the experimental time (one hour), it is considered to be in the bulk. This approach is motivated by the fact that, in experiments, active colloids rarely leave either of the walls due to self-generated hydrodynamic and chemical fields \cite{dietrich2017two,das2015boundaries,uspal2015self,popescu2018effective}. The process is repeated for $N=1000$ particles independently. We emphasize that the simulations do not include any free parameter; all values are taken from the experimental conditions \cite{SupplementalMaterial}. 

In spite of the simplicity of the model, numerical and experimental results show good agreement (see dotted lines and symbols in Fig.~\ref{fig2}(a-c)); a very pronounced negative gravitaxis takes place as the particles become, on average, density matched with the solvent, {\sl i.e.} as $F$ decreases (Fig.~\ref{fig2}(a)). Even in presence of an aligning torque, there is a critical swimming velocity, $V_{0}^{*}$, below (above) which the particles migrate preferentially downwards (upwards) (Fig.~\ref{fig2}(b-c)). $V_{0}^{*}$ is shown as blue dashed line in Figure \ref{fig3}(b) where we summarise the results by plotting the fraction of particles reaching the top surface as a function of both $V_{0}$ and $F$; negative gravitaxis (red area) is achieved using `fast' or `light' microswimmers.  

\begin{figure}[t]
	\includegraphics[scale=0.55]{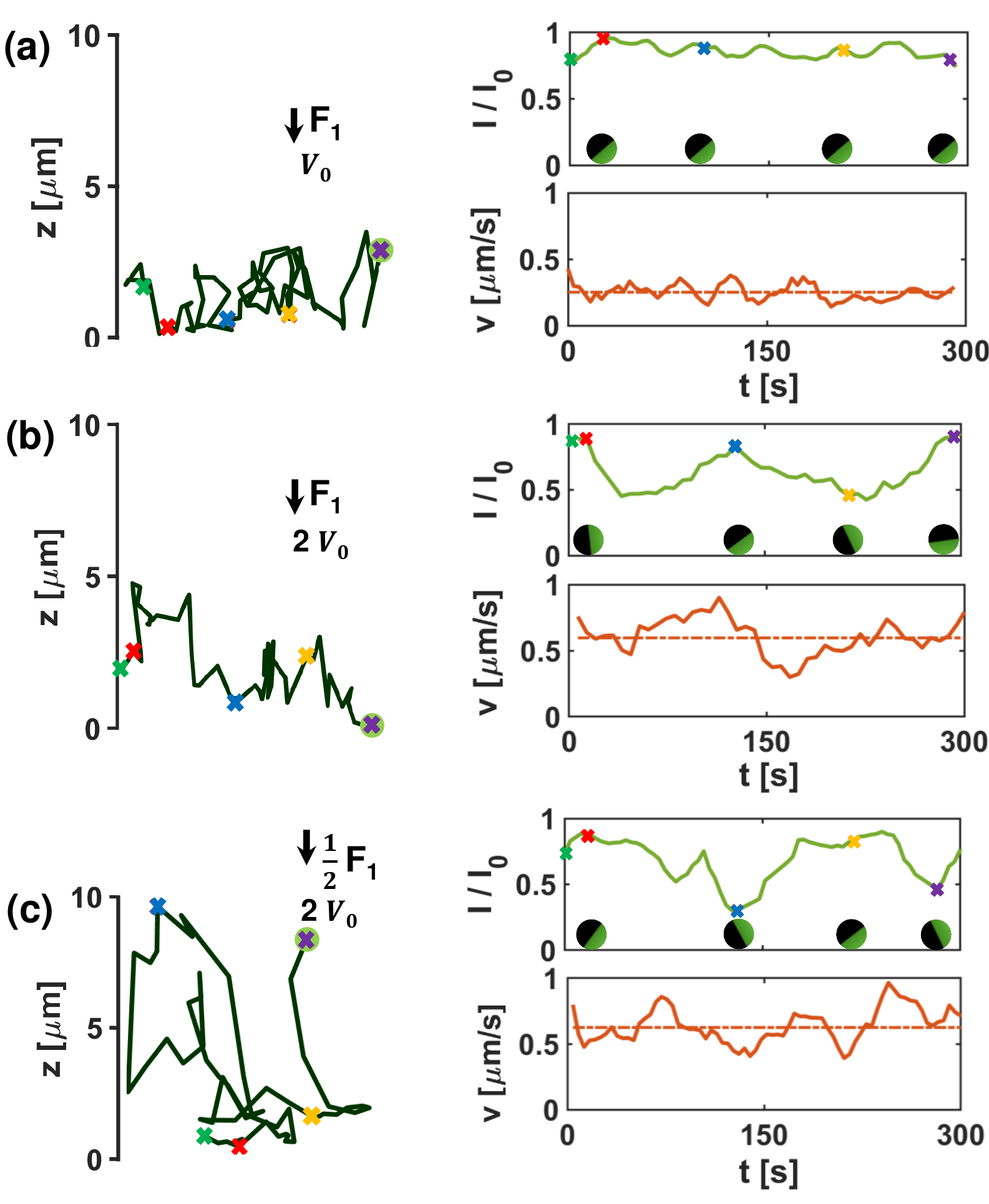}
	\caption{\label{fig4} Trajectories of active particles swimming on a glass substrate at (a) $V_0=0.3$ $\rm \mu m/s$ and (b-c) $V_0 = 0.6$ $\rm \mu m/s$ for (a-b) $F = 15.52$ $\rm fN$ and (c) $F = 9.8$ $\rm fN$. The graphs show the corresponding normalized brightness, $\rm I/I_{0}$, and instant velocity, $\rm v$, as a function of time (the horizontal dashed lines denote the average velocity of the particles). The coloured crosses in the trajectories and in the $\rm I/I_{0}$ curves mark the same times. The circles below the green curves are sketches of the active particles seen transversally in the sample, as if the $x$-axis were the bottom wall. The black and green semicircles correspond to the $\rm Pt$-cap and the uncoated hemisphere, respectively.}
\end{figure}

The buoyant weight strongly affects not only the behaviour of active particles in bulk fluids, but also their motion once they reach one of the flat walls. Existing experiments of catalytic colloids in $\rm H_2O$-$\rm H_2O_2$ suggest that the active motion near a flat wall is two-dimensional both in terms of translation and rotation; the particles swim at constant distance from the surface retaining a fixed angle between the orientation vector ({\sl i.e.} the vector linking the poles of the two hemispheres) and the $z$-axis \cite{dietrich2017two,das2015boundaries,ketzetzi2020slip,ketzetzi2020diffusion,das2020floor}. Interestingly, some experiments report an orientation vector perfectly parallel to the underlying substrate \cite{das2015boundaries}, while others suggest an inclination \cite{dietrich2017two}. In the following, we briefly show that gravity significantly enriches this scenario.

We acquire $z$-stacks of active colloids swimming near the bottom wall at $0.25$ frames per second using an inverted confocal microscope (Nikon, objective 60x Oil, NA=1.4) (one example is shown in Fig.~\ref{fig1}(c), green line). Figures \ref{fig4}(a) and \ref{fig4}(b) show the (3D) trajectories (extracted using a Python tracking package \cite{allan_daniel_2016_60550}) in $\rm H_2O$-$\rm H_2O_2$ mixtures ($F=15.52$ $\rm fN$) at $V_0=0.3$ $\rm \mu m/s$ and $V_0=0.6$ $\rm \mu m/s$. Instead, the particle in Fig.~\ref{fig4}(c) has the same swimming velocity as the one in Fig.~\ref{fig4}(b), but about half its buoyant weight ($F=9.8$ $\rm fN$). In all situations, the particles are confined near the substrate but undergo an oscillatory motion in the $z$-direction, even reaching maximum distances from the underlying wall larger than $7R$ (Fig.~\ref{fig4}(c)).

From the trajectories, we calculate the instant velocity $\rm v$ in the $xy$-plane as the distance travelled by the particle between two consecutive frames (Fig.~\ref{fig4}(a-c), orange curves; note that $V_{0}$ is similar to $\langle \rm v \rangle$ but not exactly the same because $V_0$ is an ensemble average). We also measure the normalized fluorescence brightness $\rm I/I_{0}$ of each active particle (Fig.~\ref{fig4}(a-c), green curves), where $\rm I_{0}$ is the maximum brightness defined as the total intensity of a fully saturated circular region around the particle center. The particle's brightness is directly connected to its orientation, while being independent of its $z$-position; a $\rm Pt$ cap that is oriented downwards, facing the glass substrate, screens the emitted fluorescence light, thus making the particle appear darker. Since the catalytic microswimmers move in the direction of the PS hemispheres, the combination of $\rm v$ and $\rm I/I_{0}$ gives us access to the dynamics of the particles as it navigates on the substrate; small values of $\rm I/I_{0}$ correspond to particles swimming upwards (e.g. blue cross in Fig.~\ref{fig4}(c)), whereas the simultaneous observation of large $\rm I/I_{0}$ and small $\rm v$ suggests that the swimmer is temporarily pushing against the wall (e.g. red cross in Fig.~\ref{fig4}(c)). The largest values of $\rm v$ are measured when the active colloids swim parallel to the wall. 

The orientation of `slow and heavy' particles (Fig.~\ref{fig4}(a)) is predominantly fixed, although it slightly oscillates in $z$ with period $\rm T \sim 58$ $\rm s$, where $\rm T$ is extracted from the frequency of the first peak of the power spectral density of the fluorescence intensity signal (see Supplementary Material \cite{SupplementalMaterial}. The periodic reorientation leads to an oscillatory motion on top of the wall, which has been attributed to a repulsive potential between the flat confinement and the particle \cite{bayati2019dynamics}. $\rm T$ becomes much larger for faster (Fig.~\ref{fig4}(b), $\rm T \sim 128$ $\rm s$) and lighter (Fig.~\ref{fig4}(c), $\rm T \sim 148$ $\rm s$) colloids, suggesting that gravity plays an important role in the rotational and translational motion of active swimmers near flat walls. A possible explanation lies in the fact that wall-particle interactions are more pronounced for heavy particles swimming closer to the substrate, leading to a preferential steady orientation quasi-parallel (or tilted) with respect to the wall. On the other hand, smaller values of $F$ promote larger separation distances, thus allowing the colloids to reorient in three dimensions.

To summarize, we investigated the active motion of catalytic colloids as a function of their buoyant weight, $F$, and swimming velocity, $V_0$. Our results lead to two main conclusions. (1) Weightlessness is not easy to achieve in suspensions of patchy self-propelling particles; gravitational torques caused by the different density of the two hemispheres favour the migration towards the top boundary even when the average buoyant weight is zero. (2) The reorientation of catalytic microswimmers near flat walls -- thus their quasi two-dimensional motion -- strongly depends on $F$ and $V_0$, far beyond what has been reported to date. Further studies will address the role of friction in active matter systems; we envisage that, due to 3D reorientation, active colloids will experience a fluctuating friction coefficient as they roam on rough substrates.

\begin{acknowledgments}
 We thank Juliane Simmchen for inspiring discussions. We are also grateful to Carolina van Baalen for helping us sputter-coat the particles, and Stefan Egelhaaf and Patrick Laermann for the access to the confocal microscope. 
\end{acknowledgments}

\bibliography{references}

\end{document}


\title{Sedimentation and Levitation of Catalytic Active Colloids: Supplementary Materials}

\author{V. Carrasco-Fadanelli}
\email[]{carrasv@hhu.de}
\author{I. Buttinoni}
\affiliation{Institute of Experimental Colloidal Physics, Department of Physics, Heinrich-Heine-Universität Düsseldorf, 40225 Düsseldorf, Germany}

\maketitle

\section{Supplementary Material}

\subsection{Methods}

Monolayers of polystyrene spherical microparticles (radius $R = 1.4 \ \mu \rm m$, ${{{\rm{SO}}}_{4}}^{-}$ functional groups, Microparticles GmbH) are assembled by drying a droplet of a diluted suspension onto a pre-cleaned microscope slide (Thermo Scientific). After, $4$ $\rm nm$ of platinum are sputtered onto the monolayer with $90^{\circ}$ glancing angle to coat one hemisphere of the colloids. The particles are later detached from the glass substrate via sonication in milliQ water.

A solution with milliQ water, heavy water ($\rm D_2O$) and stabilized hydrogen peroxide ($\rm H_{2}O_{2}$, $30\%$ v/v, Sigma Aldrich) is prepared at different concentrations and $\rm Pt$-coated particles are added to the solution and agitated. The cell (Hellma Analytics, Flow cell 137-QS, Suprasil quartz glass, $1$ $\rm mm$ H $\rm \times$ $35$ $\rm mm$ L $\rm \times$ $9$ $\rm mm$ W) is filled with the mixture using a pipette and the out- and inlets are sealed with parafilm. The active particles are observed right after deposition with an inverted fluorescence microscope (Olympus IX73, $20 \rm x$ and $10 \rm x$ objectives) and images are recorded at $5$ frames per second by means of a CMOS camera to extract the $xy$-trajectories. We repeat the measurements for each sample after approximately one hour of deposition to ensure that the average swimming velocity is still the same after the waiting time. 

\begin{figure}[h]
\begin{center}
 \includegraphics[scale=0.55]{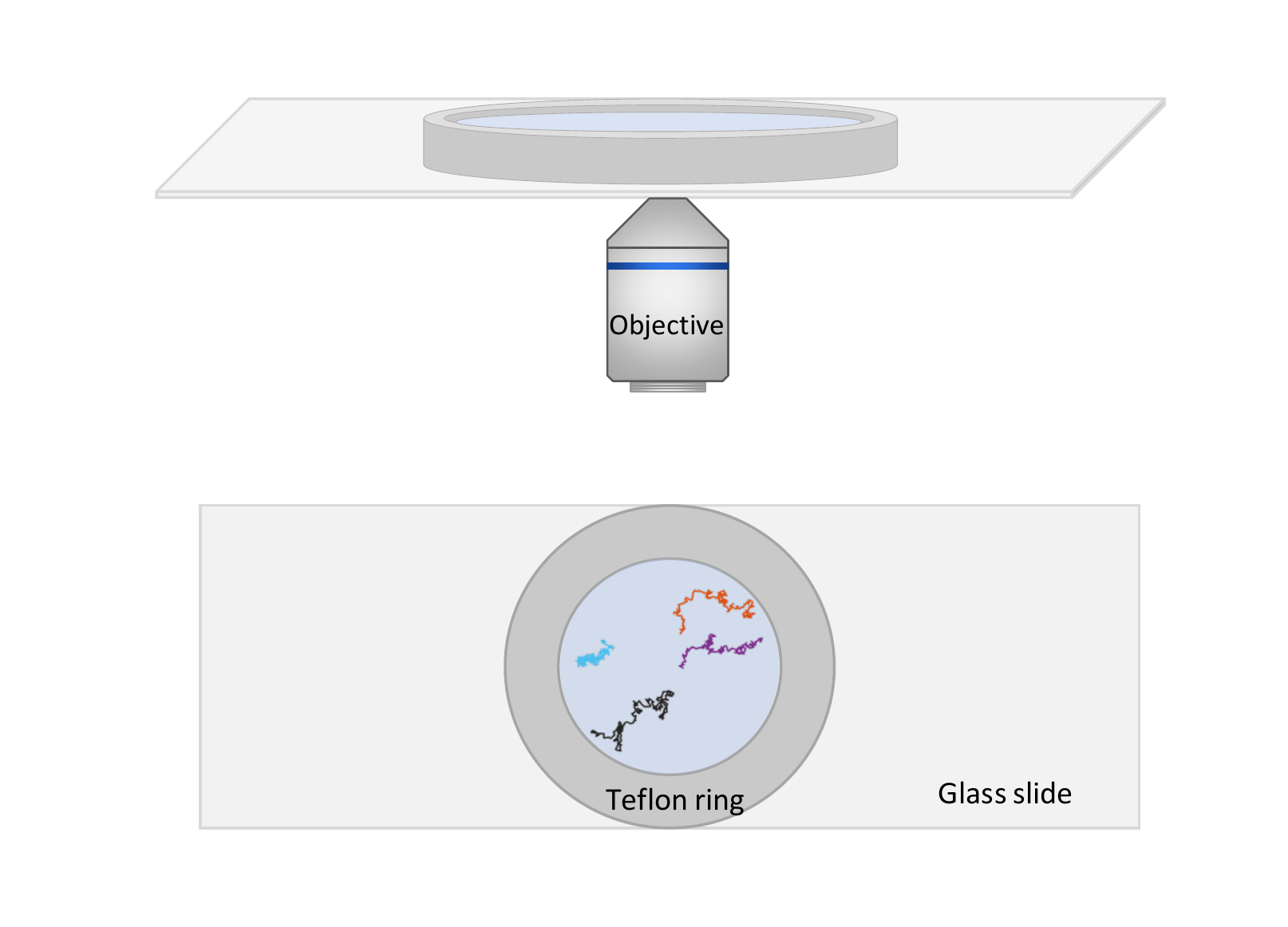}
 \caption{\label{Fig1supp} Cell used for confocal experiments.}
 \end{center}
\end{figure}

For the acquisition of three-dimensional trajectories, we use an inverted confocal microscope (Nikon, objective  $60\rm x$ Oil, $\rm NA=1.4$) to image the sample. Here, the cell consists of a Teflon ring glued on a microscope cover slide (Fig.~S\ref{Fig1supp}) and z-stacks of the particles near the bottom substrate are taken at $0.25$ frames per second.

\subsection{Data analysis}
\begin{figure}
    \centering
    \includegraphics[scale=0.45]{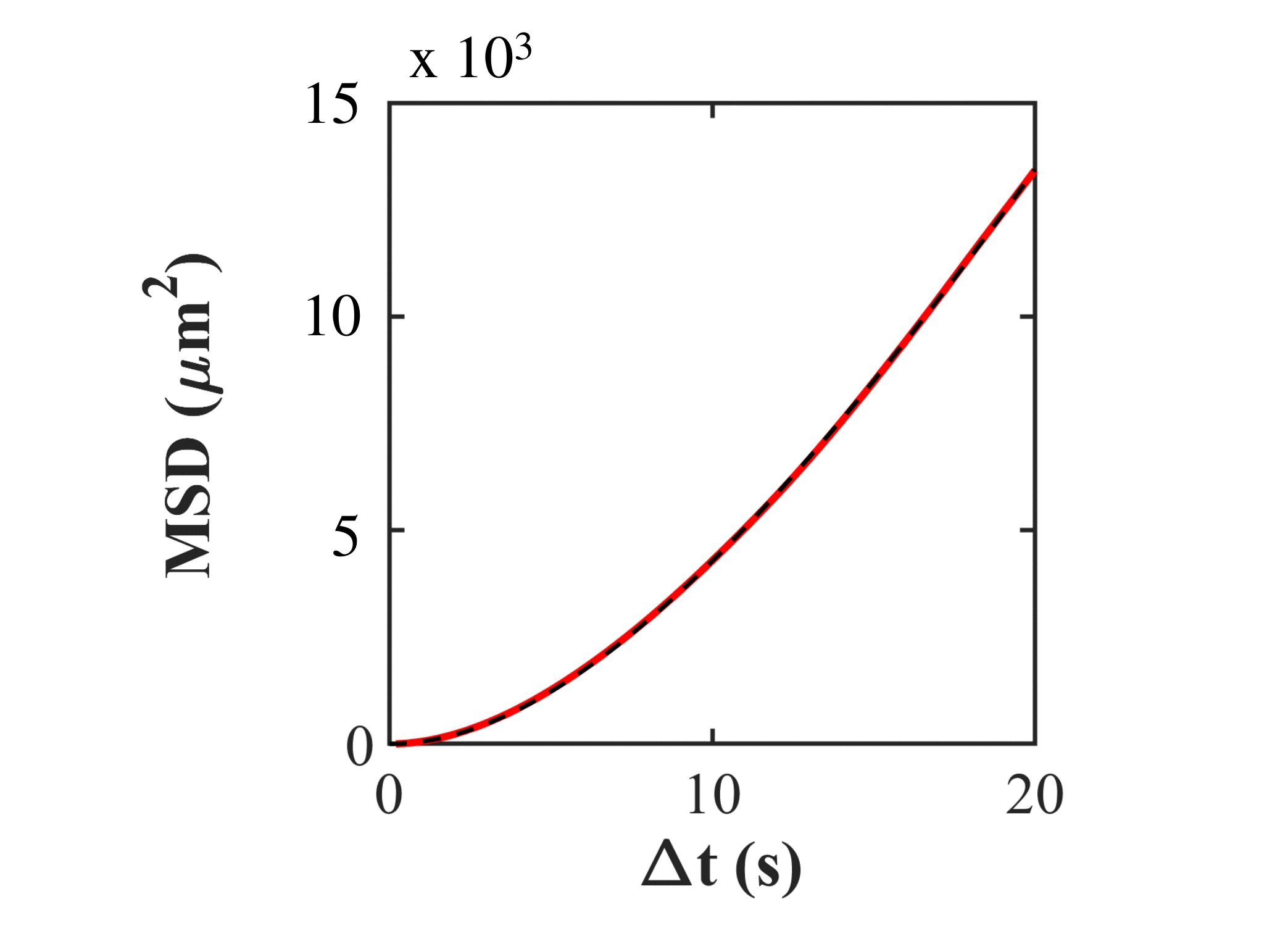}
    \caption{\label{Fig2supp} Mean squared displacement ($\rm MSD$) of a particle swimming close a planar substrate (red line). The black dashed line is the fitting curve according to Eq. (\ref{fit}).}
\end{figure}

The mean squared displacement ($\rm MSD$) of $\sim 15$ particles swimming on the bottom substrate is calculated as

\begin{equation}
    {\rm MSD}(\Delta t) = \langle (\textbf{r}(t +\Delta t) - \textbf{r}(t) )^{2} \rangle,
\end{equation}

\noindent where $\textbf{r}(t)$ and $\textbf{r}(t+\Delta t)$ are the particle's positions at $t$ and $t+\Delta t$, respectively. The average swimming velocity $V_{0}$ is extracted from the short-time fitting of the $\rm MSD$, 

\begin{equation}\label{fit}
   {\rm MSD}(\Delta t) = 4D_{0}\Delta t + V_0^{2} \Delta t^{2},
\end{equation}

\noindent where the translational diffusion coefficient $D_{0}$ is evaluated without any propulsion (no $\rm H_{2}O_{2}$). An example is given in Fig.~S\ref{Fig2supp}; the red line shows the $\rm MSD$ and the black dashed line is the fitting curve. The swimming velocity obtained using the $\rm MSD$ is very similar to the one evaluated as an average of the instant velocities.

\begin{figure}[H]
\begin{center}
 \includegraphics[scale=0.45]{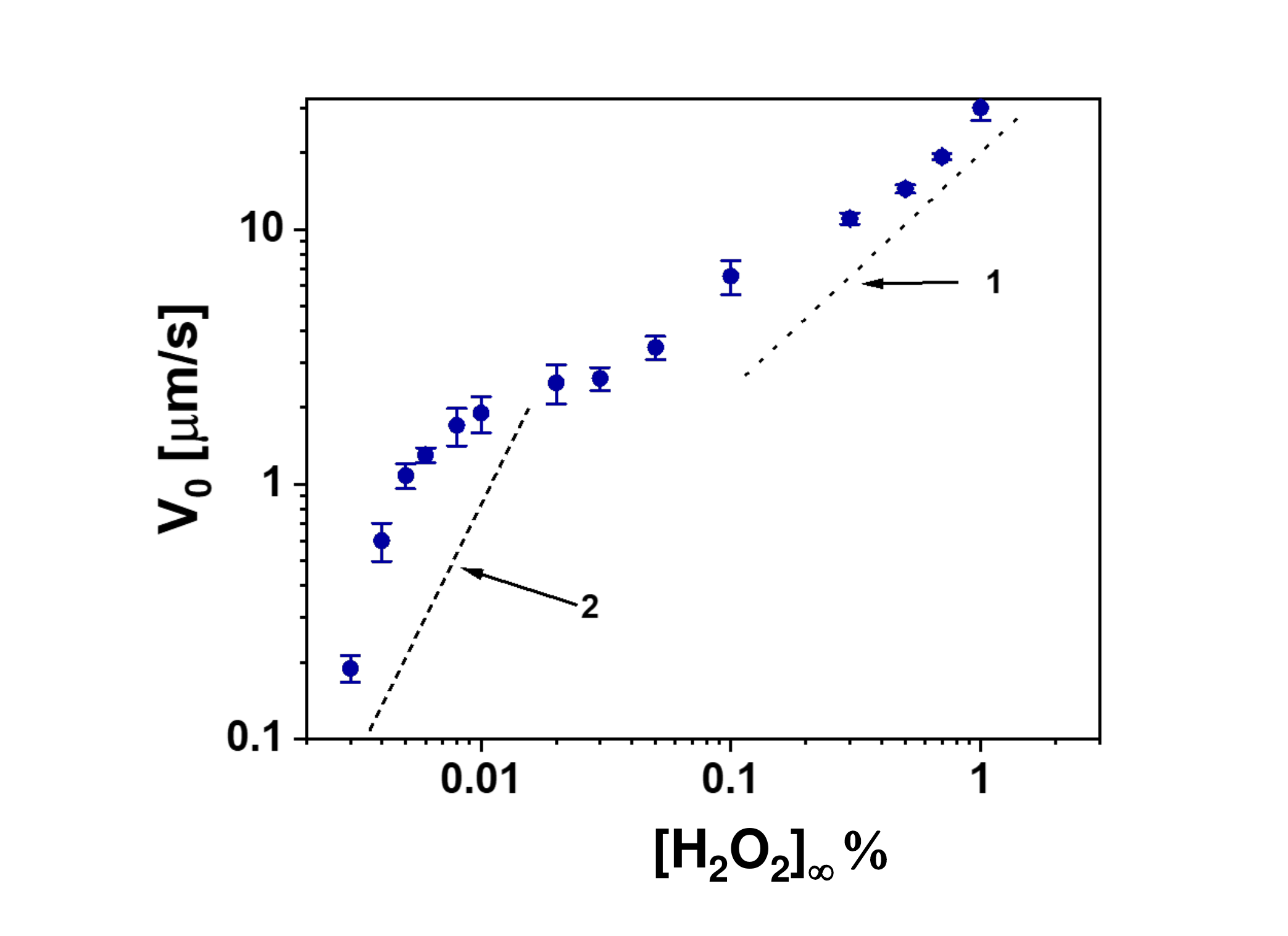}
 \caption{\label{Fig3supp} Log-log calibration curve of $V_{0}$ as a function of $[\rm H_{2}O_{2}]_{\infty}$ for $\rm [D_{2}O]=0$ $\%$. We can observe a linear growth in $\rm V_{0}$ for $[\rm H_{2}O_{2}]_{\infty} > 0.1$ $\%$. For smaller values of $[\rm H_{2}O_{2}]_{\infty}$, the behaviour of $\rm V_{0}$ is more complex. The dotted lines serve as a visual reference for a linear (dotted line 1) and quadratic behaviour (dotted line 2).}
 \end{center}
\end{figure}

Fig.~S\ref{Fig3supp} shows the calibration curve of $\rm V_{0}$ as a function of $[\rm H_{2}O_{2}]_{\infty}$ in the absence of $\rm D_2O$. For high values of $[\rm H_{2}O_{2}]_{\infty}$, $\rm V_{0}$ grows linearly as reported in the existing literature. However, for smaller values of fuel concentration, the dependency is highly nonlinear. 

\subsection{Numerical Simulations}

The following overdamped Langevin equations describe the local position $\textbf{r}=(x,z)$ of a microdisk that self-propels in a fluid of viscosity $\eta$ at velocity $V_{0}$, which is constant in magnitude but has an orientation that fluctuates in time.

\begin{equation}\label{Lang1}
  \frac{d\textbf{r}}{d \rm t}=\rm V_{0}\hat{\textbf{u}}(t) + \sqrt{2\rm D_{0}}\textbf{W}(t),	  
\end{equation}

\noindent where $D_{0}$ is the translational diffusion coefficient, $\textbf{W}(t)$ is a stochastic processes with average 0 and standard deviation 1, and $\hat{\textbf{u}}(t)=({\rm cos}(\theta(t), {\rm sin}(\theta(t))$ is the unit vector of the swimming velocity $V_0$ such that

\begin{equation}\label{Lang2}
	\frac{d\theta}{d  t}= \sqrt{2D_{\rm R}}\textbf{W}_{\theta}(t).	  
\end{equation}

\begin{figure}[H]
\begin{center}
 \includegraphics[scale=0.45]{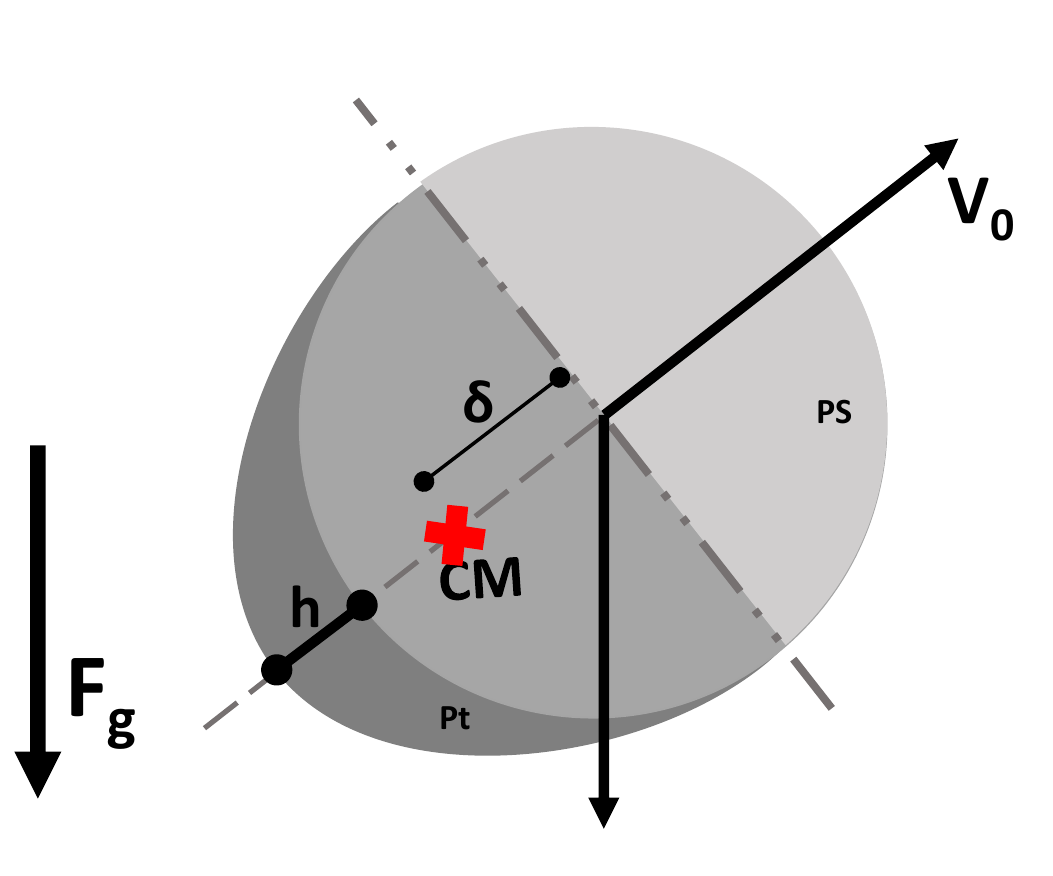}
 \caption{Center of mass (CM) of a polystyrene particle with a solid half-ellipsoidal cap of Pt (dark gray). The maximum thickness, $\rm h$, of the cap is reached at the particle's pole.}\label{Fig4supp}
 \end{center}
\end{figure}

The experimental results are reproduced by adding a gravitational force to Eq. (\ref{Lang1}) and a gravitational torque to Eq. (\ref{Lang2}), {\sl i.e.}

\begin{equation}\label{Lang1_2}
	\frac{d\textbf{r}}{d t}= V_{0}\hat{\textbf{u}}(t) - \frac{F}{6 \pi \eta R} \textbf{u}_{z} + \sqrt{2D_{0}}\textbf{W}(t),	  
\end{equation}

\begin{equation}\label{Lang2_2}
	\frac{d\theta}{d t}= \frac{\bm{\Omega}}{8 \pi \eta R^3} + \sqrt{2D_{R}}\textbf{W}_{\theta}(t),	  
\end{equation}

\noindent where $\textbf{F}$ (buoyant weight) and $\bm{\Omega}$ (gravitational torque of magnitude $\Omega=\rm F\delta \sin(\theta)$, where $\delta$ is the distance between the center of geometry and the center of mass, see Fig.~S\ref{Fig4supp}) are also defined in the text. The torque originates because of the presence of a $\rm Pt$-cap on the particle. The shape the cap is considered to be an solid half ellipsoid, in which the thinner part of the layer is in the equator of the particle and thickens towards the top, where the layer reaches a maximum thickness, $\rm h = 4$ $\rm nm$, as shown in Fig.~S\ref{Fig4supp}.

The particle's center of mass ($ \rm CM$) is therefore displaced by a distance $\rm \delta$ from the geometrical center, 

\begin{equation}
	\delta=\frac{3(R+h)}{8R}(R+h)w-\frac{3}{8}wR,
\end{equation}

\noindent where $w$ is the weight of half-disk made of platinum.

\subsection{Sedimentation Length and Maximum Torque} 

In this Section, we report the values of the sedimentation length, $l_{\rm g}= k_{\rm B}T_{\rm eff}/F$, and the maximum torque, $\Omega_{\rm max}= F\delta$, of catalytic active particles, for the different buoyant weights investigated in the manuscript.

\begin{figure}[H]
	\begin{center}
		\includegraphics[scale=0.8]{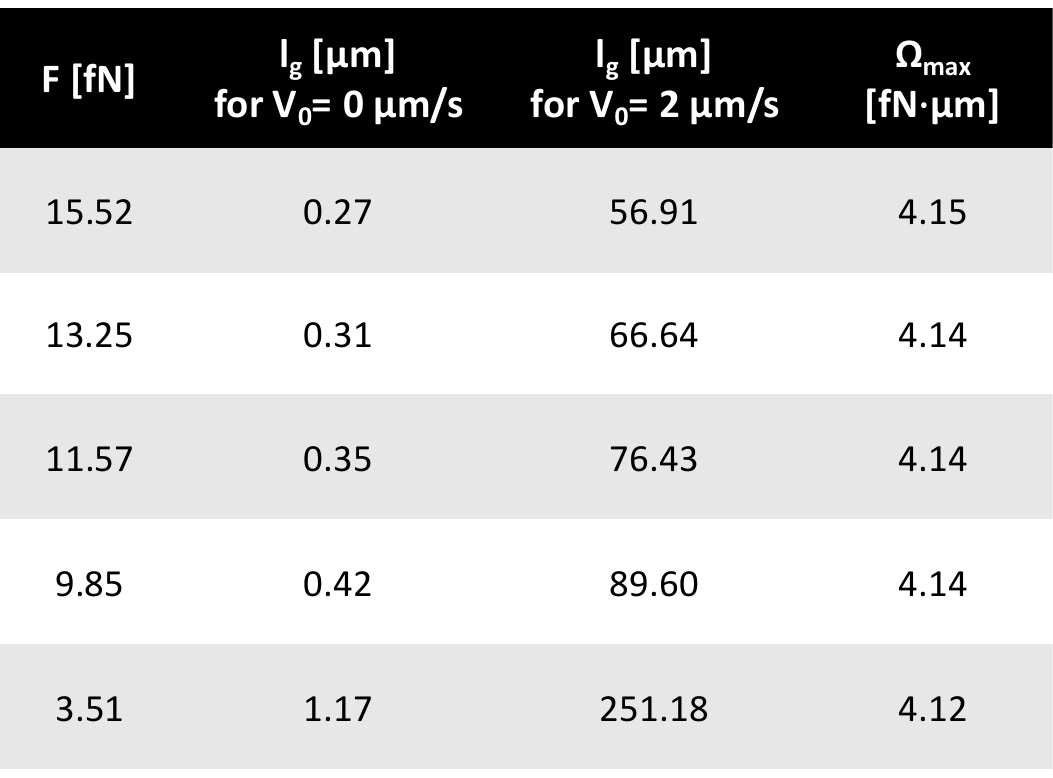}
		\caption{\label{Fig5supp} Sedimentation length, $l_{\rm g}$, and maximum torque, $\Omega_{\rm max}$, corresponding to different buoyant weights $\rm F$ and swimming velocities $V_0$.}
	\end{center}
\end{figure}

\subsection{Power spectra of the intensity}

\begin{figure}[H]
	\begin{center}
		\includegraphics[scale=0.55]{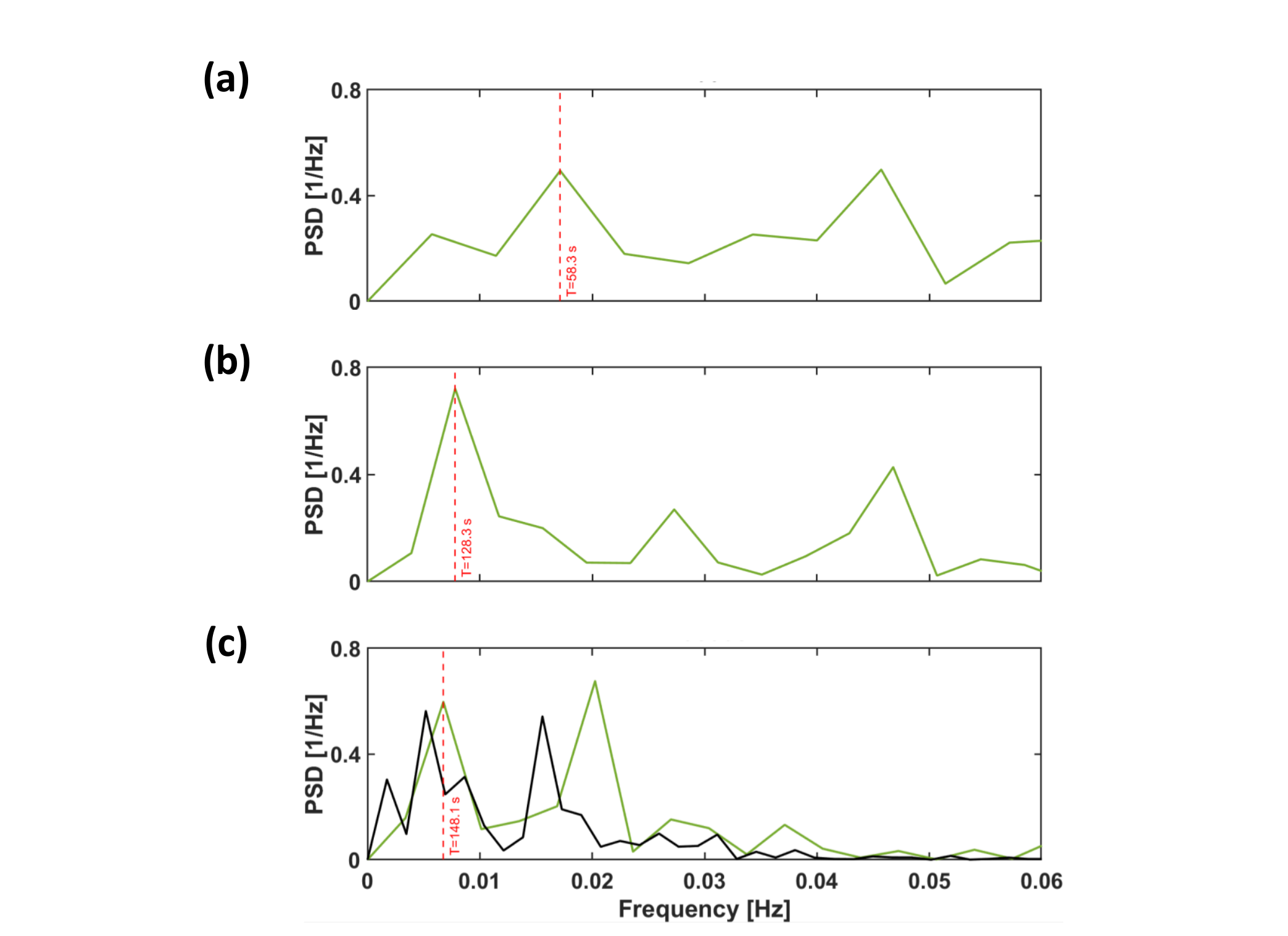}
		\caption{\label{Fig6sup} Power spectral density of intensity (green lines) and height (black lines) for the same cases as in Fig. 4 in the main text: (a) $\rm v=0.3 \ \mu m/s$ and $\rm F=15.52 \ fN$, (b)  $\rm v=0.6 \ \mu m/s$ and $\rm F=15.52 \ fN$, (c)  $\rm v=0.6 \ \mu m/s$ and $\rm F=9.8 \ fN$. The red dotted lines show where the first peak for the intensity apppears.}
	\end{center}
\end{figure}

The power spectral densities (PSD) of the fluorescence intensity of the particles (obtained by confocal microscopy) is shown in Fig.~S\ref{Fig6sup} (green lines) for the three cases shown in Fig. 4 of the main text. A peak at low frequencies is always observed, indicating that the particles undergo oscillatory reorientation near the flat wall. The first peak shifts to lower frequencies as $V_0$ is increased (Fig.~S\ref{Fig6sup}(b)) or $F$ decreases (Fig.~S\ref{Fig6sup}(c)). The periodic reorientation leads to an oscillatory motion in $z$, as shown by the PSD of the $z$-coordinates of the trajectory (Fig.~S\ref{Fig6sup}(c), black line).

